\documentclass[conference]{IEEEtran}
\usepackage{amsmath}
\usepackage{amsfonts}
\usepackage{bbding}
\usepackage{amssymb}
\usepackage{array}
\usepackage{subfigure}

\usepackage{graphicx}
\usepackage{subfigure}
\usepackage[named]{algo}
\usepackage{algorithmic}
\usepackage{psfrag}
\usepackage{stfloats}
\usepackage[compress]{cite}
\makeatletter
\renewcommand{\citepunct}{,\penalty\@m\hskip.13emplus.1emminus.1em}
\renewcommand{\citedash}{\hbox{--}\penalty\@m}
\makeatother
\usepackage{setspace}
\usepackage{color}
\allowdisplaybreaks

\usepackage{amsthm}
\usepackage{stfloats}

\usepackage{hyperref}
\usepackage{bm}

\begin{document}
\title{Energy Efficient Design for Tactile Internet}

\author{
\IEEEauthorblockN{{Changyang She and Chenyang Yang}} \vspace{0.0cm}
\IEEEauthorblockA{School of Electronics and Information
Engineering, Beihang University, Beijing, China\\
Email:  \{cyshe,cyyang\}@buaa.edu.cn} }

\maketitle
\begin{abstract}
Ensuring the ultra-low end-to-end latency and ultra-high reliability required by tactile internet is challenging. This is especially true when the stringent Quality-of-Service (QoS) requirement is expected to be satisfied not at the cost of significantly reducing spectral efficiency and energy efficiency (EE). In this paper, we study how to maximize the EE for tactile internet under the stringent QoS constraint, where both queueing delay and transmission delay are taken into account. We first validate that the upper bound of queueing delay violation probability derived from the effective bandwidth can be used to characterize the queueing delay violation probability in the short delay regime for Poisson arrival process. However, the upper bound is not tight for short delay, which leads to conservative designs and hence leads to wasting energy. To avoid this, we optimize resource allocation that depends on the queue state information and channel state information. Analytical results show that with a large number of transmit antennas the EE achieved by the proposed policy approaches to the EE limit achieved for infinite delay bound, which implies that the policy does not lead to any EE loss. Simulation and numerical results show that even for not-so-large number of antennas, the EE achieved by the proposed policy is still close to the EE limit.
\end{abstract}

\section{Introduction}
{Tactile internet} enables unprecedented mobile applications such as vehicle collision avoidance, mobile robots, virtual reality
and augmented reality \cite{Gerhard2014The,A2014Scenarios}, which calls for ultra-low latency (say 1 ms) and ultra-high reliability (say $99.99999$\%).
To ensure the low end-to-end (E2E) delay and high reliability for each short packet, both transmission and process delay and queueing delay should be bounded with small violation probability  \cite{Changyang2016TVC}, and the delay spent in the backbone network should be controlled by updating the  network architectures. By introducing short frame structure, short transmit time interval (TTI)  \cite{Petteri2015A} and  using short codes such as Polar codes  \cite{Kai2014Polar}, the transmission, processing and coding delay can be reduced.

Though satisfying such a stringent quality of service (QoS) itself is rather challenging, it is not expected to be achieved at the cost of significantly reducing the spectral efficiency and energy efficiency (EE), which are important metrics for the fifth generation (5G) networks \cite{YangGR2015}
To guarantee such a stringent QoS, the resource allocation could be conservative, which may leads to a waste of energy. Moreover, to ensure the delay that may even be shorter than the channel coherence time, channel inversion power allocation is required in single-user case, which leads to unbounded transmit power. This suggests that
the EE of tactile internet systems may be low. As far as the authors known, the EE related issues has not been considered in the context of  in tactile internet.

The QoS requirement of tactile internet can be characterized by a delay bound (say, including air interface delay and queue delay) and a delay bound violation probability (say, including the queueing delay violation probability, packet loss, and drop probability).
Improving the EE under the queueing delay bound and delay bound violation probability constraint has been widely studied in existing studies, e.g, \cite{EELIU,She2015Tcom}.
Effective bandwidth and effective capacity is a powerful tool in designing resource allocation under such a statistical delay requirement \cite{EB}. However, since the distribution of queueing delay is obtained based on large deviation principle, effective bandwidth is widely believed useful only for optimizing the system with large delay requirement. It is unclear whether it can be used for design tactile internet with the short delay.

In this paper, we study how to maximize EE by optimizing resource allocation under the QoS provision of tactile internet. We validate that the effective bandwidth can be used as a tool in the short delay regime. In fact, for the applications with ultra-low latency, an upper bound of queueing delay violation probability derived from effective bandwidth can be applied for Poisson process and the arrival processes that are more bursty than Poisson \cite{squeezing1996}. However,  the upper bound of the queueing delay violation probability is not tight, which inevitably leads to conservative design. To avoid wasting energy by the conservative designs, a queue state information (QSI) and channel state information (CSI) dependent resource allocation policy is proposed. Our analysis shows that the proposed policy is optimal in large scale antenna systems, and can achieve the EE limit obtained for the infinite delay bound. This implies that ensuring the ultra-low E2E delay and ultra-high reliability will to cause EE loss if the optimal policy is applied. As a by-product, we also derive the bandwidth and power required to guarantee the QoS. Simulation and numerical results validate our analysis and show that even with not-so-large  number of antennas, the achieved EE of the proposed policy is closed to the EE limit.

\section{System Model}
Consider a time division duplexing cellular system, where $K$ single-antenna users are served by a BS with $N_t$ antennas during successive frames. Each frame is with duration $T_f$, which consists of a downlink (DL) and an uplink (UL) transmission phase. In the UL phase, each user (i.e., a vehicle) uploads its safety messages (e.g., speed and location \cite{Mehdi2013Performance}) with short packets to the BS. In the DL phase, the BS sorts the received safety messages from the nearby users of each user, and then transmits the relevant messages to the target users.
To capture the essence of the problem, we consider frequency division multiple access to avoid the interference among multiple users.


For the tactile service, the QoS can be characterized by an E2E delay bound for each packet,  $D_{\max}$, and a packet loss/error probability, ${\varepsilon _D}$. The E2E delay is very short, say 1 ms \cite{Gerhard2014The}, which includes UL and DL transmission delay, processing and coding delay, and queueing delay in the buffer of BS. To ensure the  transmission delay ultra-low, we consider the short frame structure proposed  in \cite{Petteri2015A}, where the TTI is the same as the frame duration and  $T_f \ll D_{\max}$. Moreover, we assume that some sort of short codes can be applied such that the processing and coding delay is very low. Since the packet size is small (say less than 100 bytes), UL and DL transmission of each packet can be finished within a frame \cite{Petteri2015A}. As a consequence, the maximal queueing delay  of each packet allowed by the service is $D^q_{\max} \triangleq D_{\max}-T_f$. Denote the maximal queueing delay violation probability  allowed by the service as ${\varepsilon^q}$. Then, the  requirement imposed on the queueing delay for each packet is $(D^q_{\max}, {\varepsilon^q})$, where ${\varepsilon^q} < {\varepsilon _D}$.

\begin{figure}[htbp]
        \vspace{-0.3cm}
        \centering
        \begin{minipage}[t]{0.45\textwidth}
        \includegraphics[width=1\textwidth]{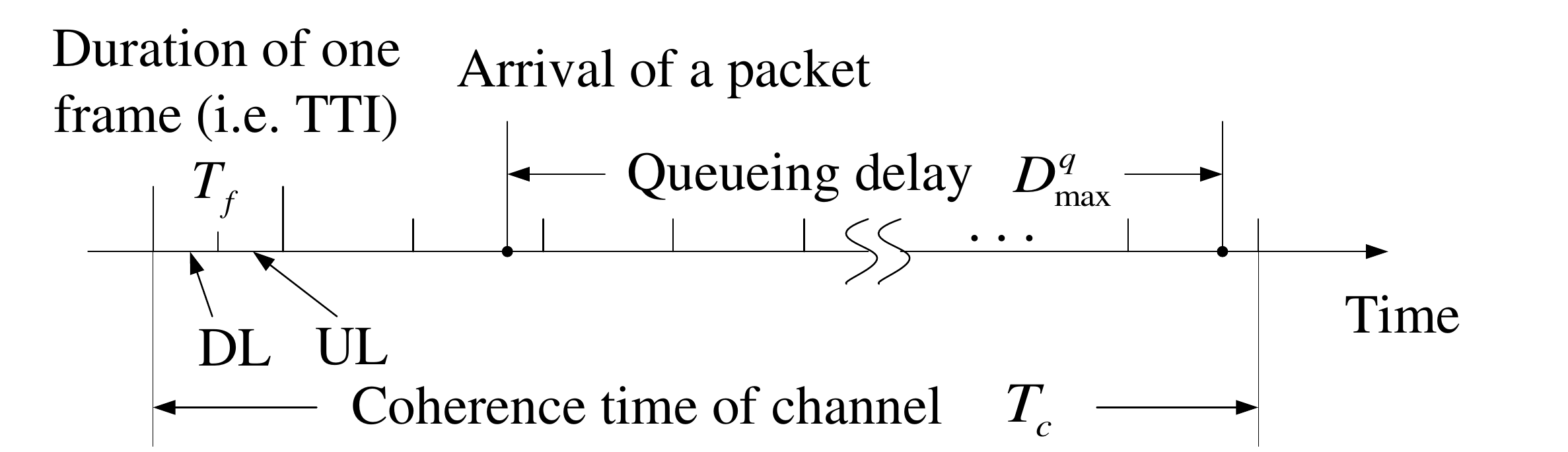}
        \end{minipage}
        \vspace{-0.2cm}
        \caption{Relation of delay bound and coherence time.}
        \label{Illustration}
        \vspace{-0.3cm}
\end{figure}

Consider block fading channel, which remains constant within each coherent interval of  duration $T_c$ and changes independently among the intervals.
For the users with velocities of 30 $\sim$ 120 km/h and the system operating in carrier frequency of 2 GHz, the channel coherence time  is around $1.1 \sim 4.5$~ms, which is larger than the queueing delay of each packet (i.e., $T_c > D^q_{\max}$), as illustrated in Fig. \ref{Illustration}. In this paper, we consider such a typical scenario of vehicular communication systems, which is more challenging than the other case with $T_c \leq D^q_{\max}$ in terms of stringent delay performance. For notational simplicity, $T_c$ is assumed to be divisible by $T_f$.

Due to the low rate requirement of each user, the bandwidth allocated to each user is usually less than the coherent bandwidth of the channel. Hence, it is reasonable to assume flat fading. Denote the average channel gain and channel vector of the $k$th user in a certain coherence interval as $\alpha_k$ and ${\bf{h}}_k \in {\mathbb{C}}^{N_t \times 1}$, whose elements are independent and identically Gaussian distributed with zero mean and unit variance. When $\alpha_k$ and  ${\bf{h}}_k$ are perfectly known at the BS and the user, the maximal number of packets that \emph{can be} transmitted to the $k$th user in the $n$th frame is given by
\begin{align}
s_k(n) = {\frac{{\Phi} T_D W_k(n)}{u } \log_2\left[1+\frac{\alpha_k P^t_k(n)g_k}{ N_0 W_k(n)}\right]}\;
\text{(packets)}, \label{eq:sn}
\end{align}
where $u$ is the size of each packet, $P^t_k(n)$ and $W_k(n)$ are respectively the transmit power and bandwidth allocated to the $k$th user according to its queue length in the $n$th frame,  $T_D$ is the duration of DL transmission phase, $N_0$ is the single-sided noise spectral density, $g_k = {\bf{h}}_k^H{\bf{h}}_k$ is instantaneous channel gain, $[ \cdot ]^H$ denotes the conjugate transpose, and ${\Phi} \in (0,1]$ is the gap between channel capacity and data rate achieved by finite blocklength codes under given error probability ${\varepsilon ^c}$ \cite{Yury2010Channel}.

\begin{figure}[htbp]
        \vspace{-0.3cm}
        \centering
        \begin{minipage}[t]{0.45\textwidth}
        \includegraphics[width=1\textwidth]{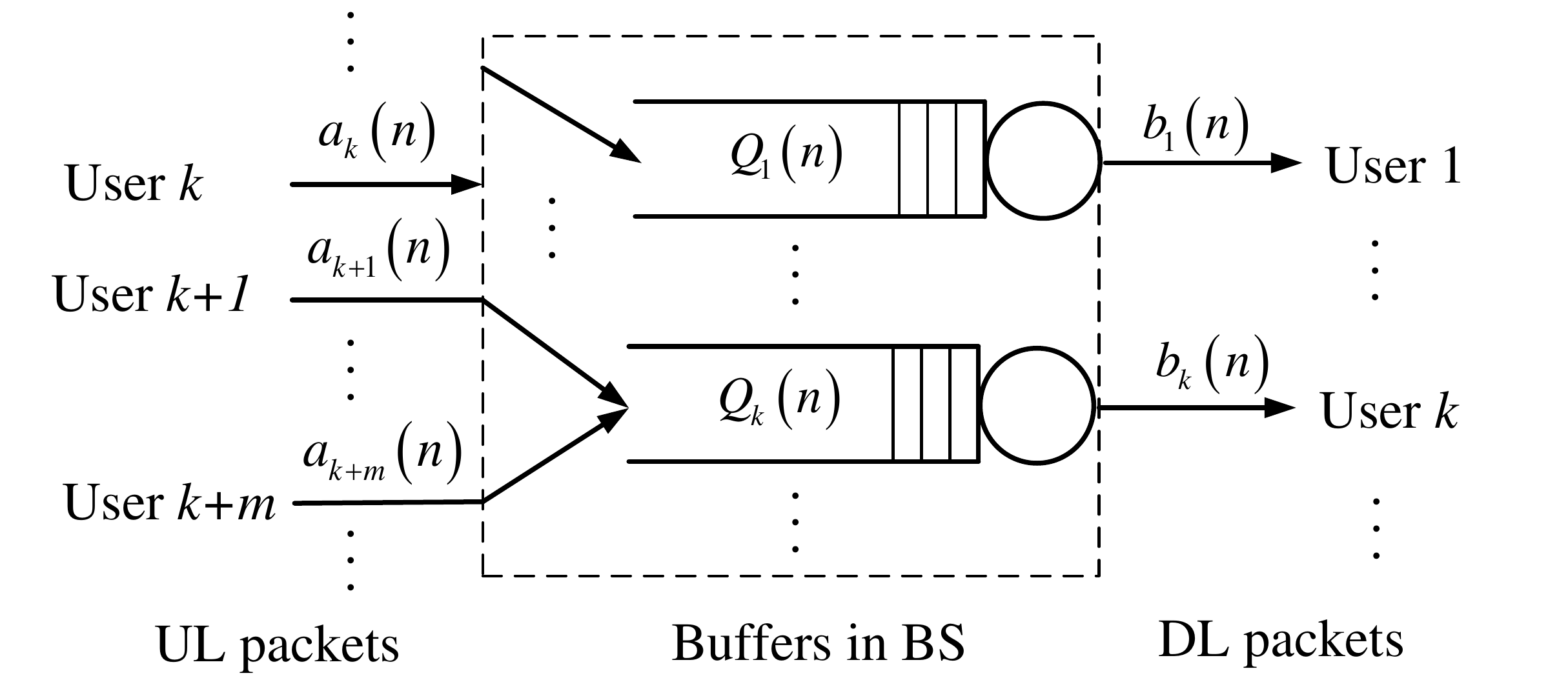}
        \end{minipage}
        \vspace{-0.3cm}
        \caption{Queueing model at the BS.}
        \label{fig:model}
        \vspace{-0.3cm}
\end{figure}

In the $n$th frame, the $k$th user requests the packets uploaded from its nearby users, whose indices constitute a set ${\mathcal{A}}_k$ with cardinality $\left|{\mathcal{A}}_k\right|$. As illustrated in Fig. \ref{fig:model}, the index set of the nearby users of the $k$th user is ${\mathcal{A}}_k = \{k+1,...,k+m\}$.
Then, the number of packets waited in the queue for the $k$th user at the beginning of the $(n+1)$th frame can be expressed as
\begin{align}
Q_k\left( {n + 1} \right) = \max \left\{ {Q_k\left( n \right) - s_k\left( n \right)  },0 \right\}+ \sum\limits_{i \in {\mathcal{A}}_k} {{a_i}\left( n \right)}, \label{eq:queue}
\end{align}
where $a_i\left( n \right)$, $i \in {\mathcal{A}}_k$ is the number of packets uploaded to the BS from the $i$th nearby user of the $k$th user.

We  consider the scenario that the inter-arrival time between packets could be shorter than $D^q_{\max}$ (otherwise the queueing delay is zero), which happens when the packets for a target user are randomly uploaded from multiple nearby users, i.e., $\left|{\mathcal{A}}_k\right|>1$. Intuitively, such a scenario seems to occur with a low probability. However, to ensure the ultra-high reliability of ${\varepsilon _D}=0.001$\%$\sim$$0.00001$\% \cite{Gerhard2014The,A2014Scenarios}, the scenario of none-zero queueing delay is not negligible.

Denote the number of packets departed from the $k$th queue in the $n$th frame as $b_k(n)$. If all the packets in the queue can be successfully transmitted in the $n$th frame, then $b_k(n) = Q_k(n)$. Otherwise, $b_k(n)=s_k(n)$. Hence, we have
\begin{align}
b_k(n) = \min \left\{Q_k\left( n \right), s_k\left( n \right)\right\}. \label{eq:bn}
\end{align}

Considering \eqref{eq:queue} and \eqref{eq:bn}, we can show after some regular derivations that the queue length evolves as follows
\begin{align}
Q_k\left( {n + 1} \right) - Q_k\left( n \right) = \sum\limits_{i \in {\mathcal{A}}_k} {{a_i}\left( n \right)} - b_k(n), \label{eq:DeltaQ}
\end{align}
which implies that the queueing delay can be controlled by adjusting the departure process.

\section{Ensuring the Queueing Delay Requiremet}
In this section, we employ \emph{effective bandwidth} to represent the QoS constraint imposed on the queueing delay. Then, we present a M/D/1 queueing model, with which we validate that effective bandwidth can be applied in the short delay regime.

\subsection{Representing QoS Constraint with Effective Bandwidth}
The aggregation of the packet arrival processes from the
$\left|{\mathcal{A}}_k\right|$ nearby users of the $k$th user  (i.e., $\sum\limits_{i \in {\mathcal{A}}_k} {{a_i}\left( n \right)}$ in \eqref{eq:queue}) can be modelled as a Poisson process \cite{Mehdi2013Performance}. For a Poisson arrival process, the effective bandwidth is \cite{Changyang2016TVC}
\begin{align}
E^{B}_k(\theta_k) = \frac{{{\lambda _k}}}{{T_f{\theta _k}}}\left( {{e^{{\theta _k}}} - 1} \right)\; \text{(packets/s)},\label{eq:EBPoisson}
\end{align}
where $\theta_k$ is the QoS exponent, $\lambda_k$ is the average number of packets arrived at the $k$th queue during one frame, which is identical for all frames.

When the $k$th user is served with a constant rate equal to $E^{B}_k(\theta_k)$, the steady state queueing delay violation probability can be approximated as \cite{EC}
\begin{align}
\Pr\{D_k(\infty) > D^q_{\max}\} \approx \eta_k \exp\{-\theta_k E_k^B(\theta_k) D^q_{\max}\},\label{eq:apporxD}
\end{align}
where $\eta_k$ is the buffer non-empty probability, and the approximation is accurate when $D^q_{\max} \to \infty$  \cite{EB}.

Since $\eta_k \leq 1$, we have
\begin{align}
\Pr\{D_k(\infty) > D^q_{\max}\} \leq \exp\{-\theta_k E_k^B(\theta_k) D^q_{\max}\} \triangleq P_{D_k}^{\rm UB}. \label{eq:UB}
\end{align}
If the upper bound in \eqref{eq:UB} satisfies
\begin{align}
P_{D_k}^{\rm UB} = \exp\{-\theta_k E_k^B(\theta_k) D^q_{\max}\} = \varepsilon^q, \label{eq:delay}
\end{align}
then the QoS requirement $(D^q_{\max}, \varepsilon^q)$ can be satisfied.
We can obtain $\theta_k$ from \eqref{eq:delay} for a service with given QoS requirement and effective bandwidth,  which is a key parameter in the QoS constraint imposed on resource allocation.

Substituting \eqref{eq:EBPoisson} into \eqref{eq:delay}, we can obtain that $\theta_k = \ln \left[\frac{T_f\ln(1/\varepsilon^q)}{\lambda_k D^q_{\max}}+1 \right]$. With which \eqref{eq:EBPoisson}  can be re-expressed as,
\begin{align}
E^{B}_k(\theta_k) = \frac{\ln(1/\varepsilon^q)}{D^q_{\max} \ln \left[\frac{T_f\ln(1/\varepsilon^q)}{\lambda_k D^q_{\max}}+1\right]}\; \text{(packets/s)}.\label{eq:EBDepson}
\end{align}

To guarantee $(D^q_{\max}, \varepsilon^q)$, the minimal number of packets transmitted to the $k$th user in the $n$th frame should be a constant among frames that satisfies \cite{EB}
\begin{align}
s_k(n) = T_fE_k^B(\theta_k)\; (\text{packets}). \label{eq:QoS}
\end{align}


When the $k$th queue is served by the constant service process $\{s_k(n), n = 1,2,...\}$ that satisfies \eqref{eq:QoS}, the departure process in  \eqref{eq:bn} becomes
\begin{align}
b_k(n) = \min\{Q_k(n),T_fE^B_k(\theta_k)\}\; (\text{packets}). \label{eq:QoSbn}
\end{align}
Therefore, if the departure process  $\{b_k(n), n = 1,2,...\}$ satisfies \eqref{eq:QoSbn},
then $(D^q_{\max}, \varepsilon^q)$ can be guaranteed.

\subsection{Validating the Upper Bound $P_{D_k}^{\rm UB}$ with M/D/1 Model} As defined in \cite{EB}, the effective bandwidth is applicable for the scenario when $D^q_{\max} \to \infty$. In other words, the approximation in \eqref{eq:apporxD} is accurate when the delay bound is sufficient large. However, it is unclear when the value of $D^q_{\max}$ is large enough for an accurate approximation. One possible reason is that it is very difficult to obtain an accurate distribution of the the queue length or queueing delay.

Yet the real concern for the problem at-hand is whether the upper bound in \eqref{eq:UB} is applicable. If $P_{D_k}^{\rm UB}$ is indeed an upper bound of $\Pr\{D_k(\infty) > D^q_{\max}\}$, then a transmit policy satisfying the QoS constraint in \eqref{eq:QoS} or \eqref{eq:QoSbn} can guarantee the required QoS.

When a Poisson arrival process is served by a constant service process $\{s_k(n), n = 1,2,...\}$, the well-known M/D/1 queueing model can be applied \cite{Gross1985MD1}. For a discrete state M/D/1 queue with length as integer (i.e., the number of packets), the closed-form expression of the  queue length distribution is known. Specifically, the {complimentary cumulative distribution function} (CCDF) of the steady state queue length can be expressed as $\Pr\{Q_k(\infty) > l\} = 1 - \sum\limits_{i = 1}^l {{\pi _i}}$,
where $\pi_l = \Pr\{Q_k(\infty) = l\}$ is the probability that there are $l$ packets in the queue, which is \cite{Gross1985MD1},
\begin{align}
&\pi_0 = 1-{\gamma_k},\;\pi_1 = (1-{\gamma_k})(e^{{\gamma_k}}-1),\nonumber\\
&\pi_l = (1-{\gamma_k})\times \nonumber\\
& \left\{e^{l {\gamma_k}} +\sum\limits_{i = 1}^{l - 1} {{e^{i{\gamma_k} }}{{\left( { - 1} \right)}^{l - i}}\left[ {\frac{{{{\left( {i{\gamma_k} } \right)}^{l - i}}}}{{\left( {l - i} \right)!}} + \frac{{{{\left( {i{\gamma_k} } \right)}^{l - i - 1}}}}{{\left( {l - i - 1} \right)!}}} \right]} \right\}, \nonumber\\
&(l \geq 2),\label{eq:pdfMD1}
\end{align}
with ${\gamma_k} = \lambda_k / s_k(n)$.
For a Poisson arrival process served by a constant service process,  the
CCDF of the queueing delay can be derived from Appendix D in \cite{She2015Tcom}  as
\begin{align}
\Pr\{D_k(\infty) > T_fl/s_k(n) \} = 1 - \sum\limits_{i = 1}^l {{\pi _i}}\label{eq:CCDFD}.
\end{align}

To derive a QoS constraint for resource allocation, we need to derive the expression of $s_k(n)$ as a function of $D^q_{\max}$ and $\varepsilon^q$ by setting $l = s_k(n) D^q_{\max}/T_f$ and $1 - \sum\limits_{i = 1}^l {{\pi _i}} = \varepsilon^q$. However, the expression of $\pi_l$ in \eqref{eq:pdfMD1} is complex. Thus, the expression of $s_k(n)$ cannot be obtained
in closed-form. This indicates that the M/D/1 model is hard to be used for optimizing a transmit policy to ensure the QoS.
Nonetheless, \eqref{eq:CCDFD} can be used to validate the upper bound $P_{D_k}^{\rm UB}$  in \eqref{eq:UB} via numerical results.


\section{Energy-Efficient Resource Allocation}
The EE is the ratio of the amount of successfully transmitted data to the energy consumption \cite{earth2010}, i.e., $
EE = {{(1 - {\varepsilon _D})\sum\limits_{k = 1}^K {\sum\limits_{i \in {{\cal A}_k}}{{\mathbb{E}}\left[ {{a_i}\left( n \right)} \right]} } }}/\{{T_D{\mathbb{E}}\left[ {{P_{{\rm{tot}}}}\left( n \right)} \right]}\}$, where $P_{\rm tot}(n)$ is the total power consumed at a BS for DL transmission in the $n$th frame,
which can be modeled as \cite{Bjorn2015A}
\begin{align}
P_{\rm tot}(n) = \frac{1}{\rho}\sum\limits_{k = 1}^K{P_k^t(n)}+P^{cw}\sum\limits_{k = 1}^K {{W_k}\left( n \right)}  + P_0^c, \label{eq:Ptot}
\end{align}
where $\rho \in (0,1]$ is the power amplifier efficiency, $P^{cw}$ is the circuit power consumption per unit bandwidth, and $P_0^c$ is the circuit power that is independent of bandwidth.

Since the value of $\varepsilon _D$ is very low, the nominator is almost independent of transmit policy. Hence, maximizing the EE is equivalent to minimizing the average total power consumption.
To this end, we can minimize the instantaneous power consumption ${{P_{{\rm{tot}}}}\left( n \right)}$ by optimizing resource allocation.


\subsection{Queue Length Dependent Resource Allocation}
Recall that the service process $\{s_k(n), n = 1,2,...\}$ should be a constant among successive frames satisfying  \eqref{eq:QoS}, in order to ensure the queueing delay requirement. To support such a constant service process, it is shown from  \eqref{eq:QoSbn}  that  when $Q_k(n) < T_f E_k^B(\theta_k)$, $b_k(n) < s_k(n)$. This indicates that the number of departed packets may be less than the number of packets that \emph{can be} transmitted by the system in some frames. To save energy, i.e., avoid wasting resources of the system, we introduce a queue length dependent two-state policy: when $Q_k(n) > T_fE^B_k(\theta_k)$, $s_k(n) = T_f E^B_k(\theta_k)$, otherwise $s_k(n) = Q_k(n)$, i.e.,
\begin{align}
s_k(n) = \min\{Q_k(n), T_fE^B_k(\theta_k)\}. \label{eq:twostate}
\end{align}
Substituting \eqref{eq:twostate} into \eqref{eq:bn}, we can show that the departure process has the same form as in \eqref{eq:QoSbn}. This means that if a two-state policy satisfies \eqref{eq:twostate}, then   $(D^q_{\max}, \varepsilon^q)$ can be guaranteed.

From \eqref{eq:sn}, \eqref{eq:twostate} and \eqref{eq:Ptot}, the two-state transmit power and bandwidth allocation policy that minimizes the instantaneous total power consumption under the constraint imposed on $(D^q_{\max}, \varepsilon^q)$ can be obtained from the following problem,
\begin{align}
&\mathop {\mathop {\min }\limits_{P_k^t\left( n \right),{W_k}\left( n \right),} }\limits_{k = 1,...,K}  \sum\limits_{k = 1}^K {P_k^t(n)}  + {P^{cw}\rho}\sum\limits_{k = 1}^K {{W_k}\left( n \right)} \label{eq:minPtot}\\
\text{s.t.}\;&\frac{{\Phi {T_D}{W_k}\left( n \right)}}{u}{\log _2}\left[ {1 + \frac{{{\alpha _k}P_k^t\left( n \right){g_k}}}{{{N_0}{W_k}\left( n \right)}}} \right] \nonumber\\
&= \min\{Q_k(n), T_fE^B_k(\theta_k)\},\;
k = 1,...,K.\label{eq:QRA}\tag{\theequation a}
\end{align}
To show how much resource is required to guarantee the stringent QoS, the maximal transmit power and bandwidth constraints are not considered.

To solve the problem, we relax \eqref{eq:QRA} into inequality constraints, and refer to the new problem that minimizes \eqref{eq:minPtot} under the inequality  constraints as \emph{Problem A}. It is not hard to show that \emph{Problem A} is equivalent to the original problem. Because the left hand side of \eqref{eq:QRA} is jointly concave in $P_t(n)$ and $W_k(n)$, \emph{Problem A} is convex, which can be solved by standard tools such as the interior-point method \cite{boyd}.


\subsection{Optimality of the Two-state Policy and Required Resources}
The two-state policy is heuristic, since a policy with more than two states may give rise to lower power consumption. Nonetheless, in the sequel we show that the optimized two-state policy can maximize the EE in a large $N_t$ asymptotic. With the resulting closed-form solution, we can show how much resources are required to ensure the QoS in such a reagion. The  simulations later show that the results obtained for large value of $N_t$ also hold when $N_t$ is not so large.

\subsubsection{Minimal Average Total Power Consumed by the Two-state Policy}
When $N_t \to \infty$, \eqref{eq:sn} approaches \cite{Rusek2013Scaling}
\begin{align}
s_k(n) = \frac{{\Phi{T_D}{W_k(n)}}}{u}{\log _2}\left[{1 + \frac{{\alpha_k {N_t}P_k^t(n)}}{{N_0 {W_k(n)}}}} \right]\;
\text{(packets)}.\nonumber
\end{align}
Due to channel hardening, the small scale channel fading does not affect the service process. In this case, the QoS constraint can be obtained from \eqref{eq:QRA} by replacing $g_k$ with $N_t$. The total power minimization resource allocation problem under such a QoS constraint is refer to as \emph{Problem B}.

By analyzing the Karush-Kuhn-Tucker (KKT) conditions of \emph{Problem B} using a similar way as the proof of Proposition 2 in \cite{She2015Optimal}, we can derive that the ratio of the optimal transmit power to the optimal bandwidth allocated to each user is a constant depending on $N_t$, $P^{cw}$, $\rho$ and $\alpha_k$, i.e., $\frac{P_k^{t^*}(n)}{{W^*_k}(n)} = P^{tw}_k$. Then, we can find the optimal solution of  \emph{Problem B} as follows,
\begin{align}
&P_k^{t^*}(n) = \frac{P_k^{tw}{u}\min\{Q_k(n), T_f E^B_k(\theta_k)\}}{{\Phi{T_D}}{{\log }_2}\left( {1 + \frac{{\alpha_k {N_t}}}{{N_0}}P_k^{tw}} \right)}\label{eq:optP}, \\
&{W^*_k}(n) = \frac{{u}\min\{Q_k(n), T_f E^B_k(\theta_k)\}}{{\Phi{T_D}}{{\log }_2}\left( {1 + \frac{{\alpha_k {N_t}}}{{N_0}}P_k^{tw}} \right)}. \label{eq:optK}
\end{align}
Substituting $P_k^{t^*}(n)$ and ${W^*_k}(n)$ into \eqref{eq:Ptot}, we can obtain the minimal total power consumed by the two-state policy as
\begin{align}
P_{\rm tot}^*(n) = \sum\limits_{k = 1}^K {\frac{{\left( {\frac{{P_k^{tw}}}{\rho } + {P^{cw}}} \right)u \min\{Q_k(n), T_f E^B_k(\theta_k)\}}}{{\Phi{T_D}{{\log }_2}\left( {1 + \frac{{\alpha_k {N_t}}}{{N_0}}P_k^{tw}} \right)}}} + P_0^c. \nonumber
\end{align}

With the two-state policy,  \eqref{eq:QoSbn} can be satisfied and hence $ {\mathbb{E}}[b_k\left( n \right)] = {\mathbb{E}}\{\min\{Q_k(n), T_f E^B_k(\theta_k)\}\}$. Moreover, with the ensured QoS, for ergodic arrival and departure processes we have ${\mathbb{E}}[b_k\left( n \right)]= (1 - {\varepsilon _D}) \sum\limits_{i \in {\mathcal{A}}_k} {{\mathbb{E}}\left[ {{a_i}\left( n \right)} \right]}$. Then, ${\mathbb{E}}[P_{\rm tot}^*(n)]$ consumed by the optimal two-state policy can be rewritten as
\begin{align}
\sum\limits_{k = 1}^K {\frac{{\left( {\frac{{P_k^{tw}}}{\rho } + {P^{cw}}} \right)u (1 - {\varepsilon _D}) \sum\limits_{i \in {\mathcal{A}}_k} {{\mathbb{E}}\left[ {{a_i}\left( n \right)} \right]}}}{{\Phi{T_D}{{\log }_2}\left( {1 + \frac{{\alpha_k{N_t}}}{{N_0}}P_k^{tw}} \right)}}} + P_0^c. \label{eq:ExpPtot}
\end{align}

\subsubsection{A Lower Bound of Power Consumption}
To show that the optimized two-state policy is EE-optimal, we compare ${\mathbb{E}}[P_{\rm tot}^*(n)]$ with a lower bound achieved when $D^q_{\max} \to \infty$. The minimal average power consumption obtained for $D^q_{\max} \to \infty$ is the ultimate lower bound of those for arbitrary finite $D^q_{\max}$ requirements, and the resulting EE is the EE limit.

As shown in \cite{She2015Optimal}, to ensure the QoS with infinite delay bound, the service process only needs to satisfy
\begin{align}
s_k(n) = (1 - {\varepsilon _D})\sum\limits_{i \in {\mathcal{A}}_k} {{\mathbb{E}}\left[ {{a_i}\left( n \right)} \right]}.\label{eq:slimit}
\end{align}

The lower bound can be obtained by minimizing \eqref{eq:minPtot} under constraint \eqref{eq:slimit}. We refer to this problem as \emph{Problem C}. By analyzing the KKT conditions of \emph{Problem C}  using a similar way as the proof of Proposition 2 in \cite{She2015Optimal}, we can show that the lower bound is the same as \eqref{eq:ExpPtot}.

\subsubsection{Required Maximal Transmit Power and Bandwidth}
With the closed form solution of the optimal two-state policy, we can find the required resources to maximize the EE  with guaranteed $(D^q_{\max}, \varepsilon^q)$, in order to provide guidance for designing systems serving tactile internet. The maximal transmit power and bandwidth to achieve the EE limit with ensured QoS can be obtained respectively from,
\begin{align}
P_{\rm req}^t = \mathop {\max }\limits_{n=1,2,...} {\sum\limits_{k = 1}^K {P_k^{{t^*}}\left( n \right)} },\;{W_{\rm req}} = \mathop {\max }\limits_{n=1,2,...} {\sum\limits_{k = 1}^K {W_k^*\left( n \right)} }. \label{eq:reqPK}
\end{align}
Substituting \eqref{eq:optP} and \eqref{eq:optK} into \eqref{eq:reqPK}, further considering $\min\{Q_k(n), T_f E^B_k(\theta_k)\} \leq T_f E^B_k(\theta_k)$ and the expression of $E^B_k(\theta_k)$ in \eqref{eq:EBDepson}, we can derive that
\begin{align}
P_{\rm req}^t \leq \sum\limits_{k = 1}^K {\frac{{P_k^{tw}uT_f[\ln(1/\varepsilon^q)]/(\Phi{T_D}D^q_{\max})}}{{{{\log }_2}\left( {1 + \frac{{\alpha_k {N_t}}}{{N_0}}P_k^{tw}} \right)}{ \ln \left[\frac{T_f\ln(1/\varepsilon^q)}{\lambda_k D^q_{\max}}+1\right]}}}   \label{eq:constP},\\
{W_{\rm req }} \leq \sum\limits_{k = 1}^K {\frac{{uT_f[\ln(1/\varepsilon^q)]/(\Phi{T_D}D_{\max}^q)}}{{{{\log }_2}\left( {1 + \frac{{\alpha_k {N_t}}}{{N_0}}P_k^{tw}} \right)}{\ln \left[\frac{T_f\ln(1/\varepsilon^q)}{\lambda_k D^q_{\max} }+1\right]}}}\label{eq:constW},
\end{align}
which are nearly proportional to $1/D^q_{\max}$ and $\ln(1/\varepsilon^q)$.

\section{Simulation and Numerical Results}
In this section, we first validate our analysis, and then show the resources required to guarantee the QoS of tactile internet with simulation and numerical results.

\begin{figure}[htbp]
        \vspace{-0.3cm}
        \centering
        \begin{minipage}[t]{0.4\textwidth}
        \includegraphics[width=1\textwidth]{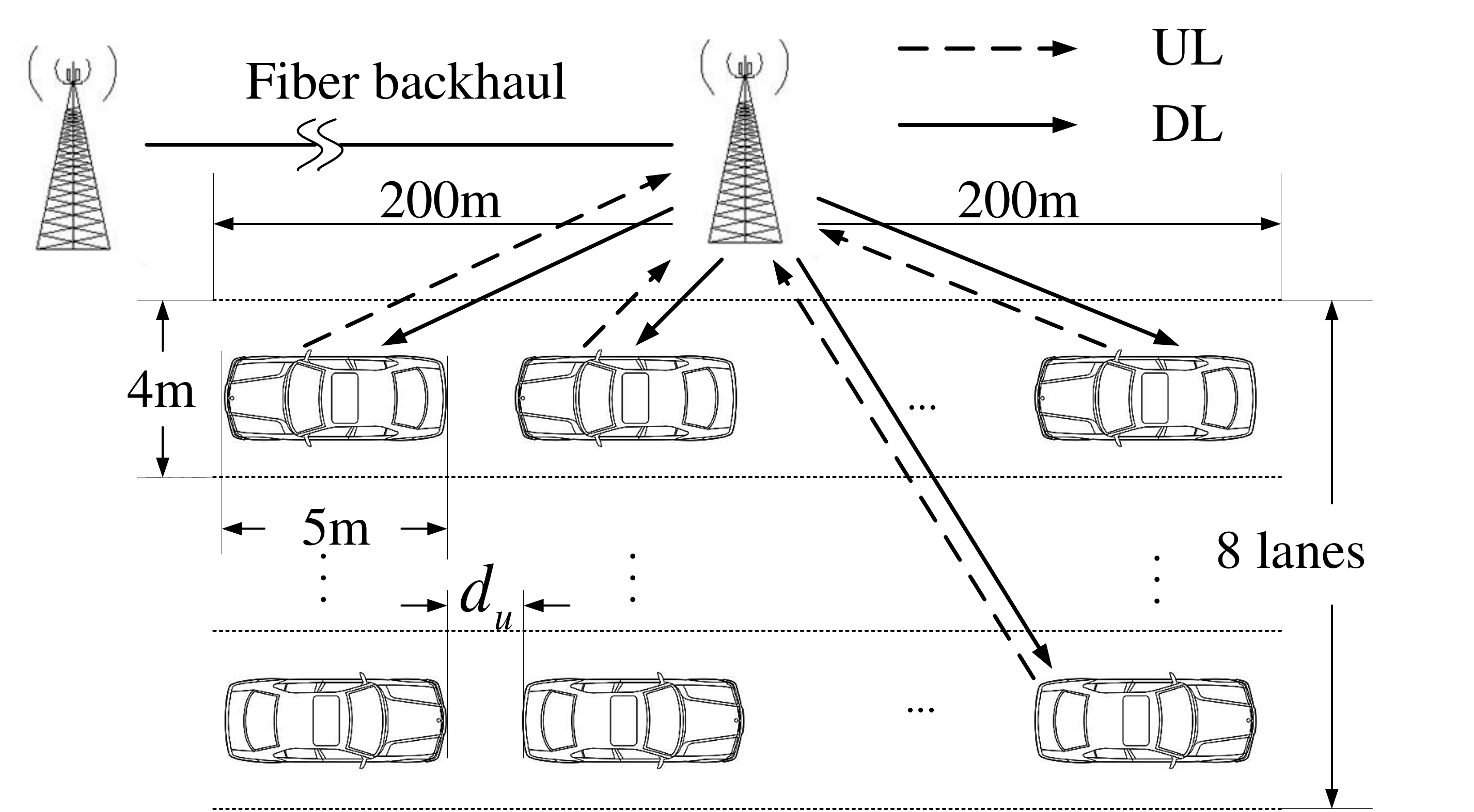}
        \end{minipage}
        \vspace{-0.2cm}
        \caption{Simulation scenario.}
        \label{fig:simulation}
        \vspace{-0.3cm}
\end{figure}

We consider an eight-lane two-direction highway scenario in urban area. The users (i.e., vehicles) uniformly located in the eight lanes are served by the roadside BSs with distance $400$ meters who are connected by fiber backhaul. The packet delay caused by fiber backhaul is around $D_B = 0.1$~ms \cite{Tony2015Delay}. The path loss model is $10\log_{10} \alpha_k = 35.3+37.6\log_{10}d_k$, where $d_k$ is the distance between a BS and the $k$th user in meters. Each vehicle requests safe messages from other vehicles with distances less than $100$~m. For the vehicles in the cell edge who request the messages from the vehicles in  adjacent cells, the BS in the adjacent cell forwards the received messages to the BS who serves the user requesting the messages, then  $D_B$ is also counted in the E2E delay, i.e., $D_{\max}^q+D_B+T_f\le D_{\max}$. Since there are other factors except the queueling delay violation lead to packet loss and error (e.g., finite blocklength channel coding), here we set $\varepsilon^q =  \varepsilon _D/2$ for simplicity. Parameters in the sequel are listed in Table I, unless otherwise specified.

\begin{table}[htbp]
\vspace{-0.3cm}
\small
\renewcommand{\arraystretch}{1.3}
\caption{List of Simulation Parameters \cite{Gerhard2014The,Yury2010Channel,Bjorn2015A}}
\begin{center}\vspace{-0.3cm}
\begin{tabular}{|p{5cm}|p{2.5cm}|}
  \hline
  End-to-end delay $D_{\max}$  & $1$~ms  \\\hline
  Reliability $1-{\varepsilon _D}$ & $99.99999$\% \\\hline
  Frame duration $T_f$ & 0.1~ms \\\hline
  Duration of DL phase $T_D$ & $0.05$~ms\\\hline
  Coherence time of channel $T_c$ & $2$~ms\\\hline
  Packet size $u$ &  $20$~bytes \\\hline
  UL average packet arrival rate &  $20$~packets/s/user \\\hline
  Data rate gap $\Phi$ & 0.9 \\\hline
  Single-sided noise spectral density $N_0$ & $-173$~dBm/Hz \\\hline
  Circuit power per unit bandwidth $P^{cw}$ & $72N_t$~mW/MHz \\\hline
  Other circuit power (e.g., cooling) $P^{c}_0$ & $136N_t$~mW \\\hline
\end{tabular}
\end{center}
\vspace{-0.4cm}
\end{table}

\begin{figure}[htbp]
        \vspace{-0.3cm}
        \centering
        \begin{minipage}[t]{0.38\textwidth}
        \includegraphics[width=1\textwidth]{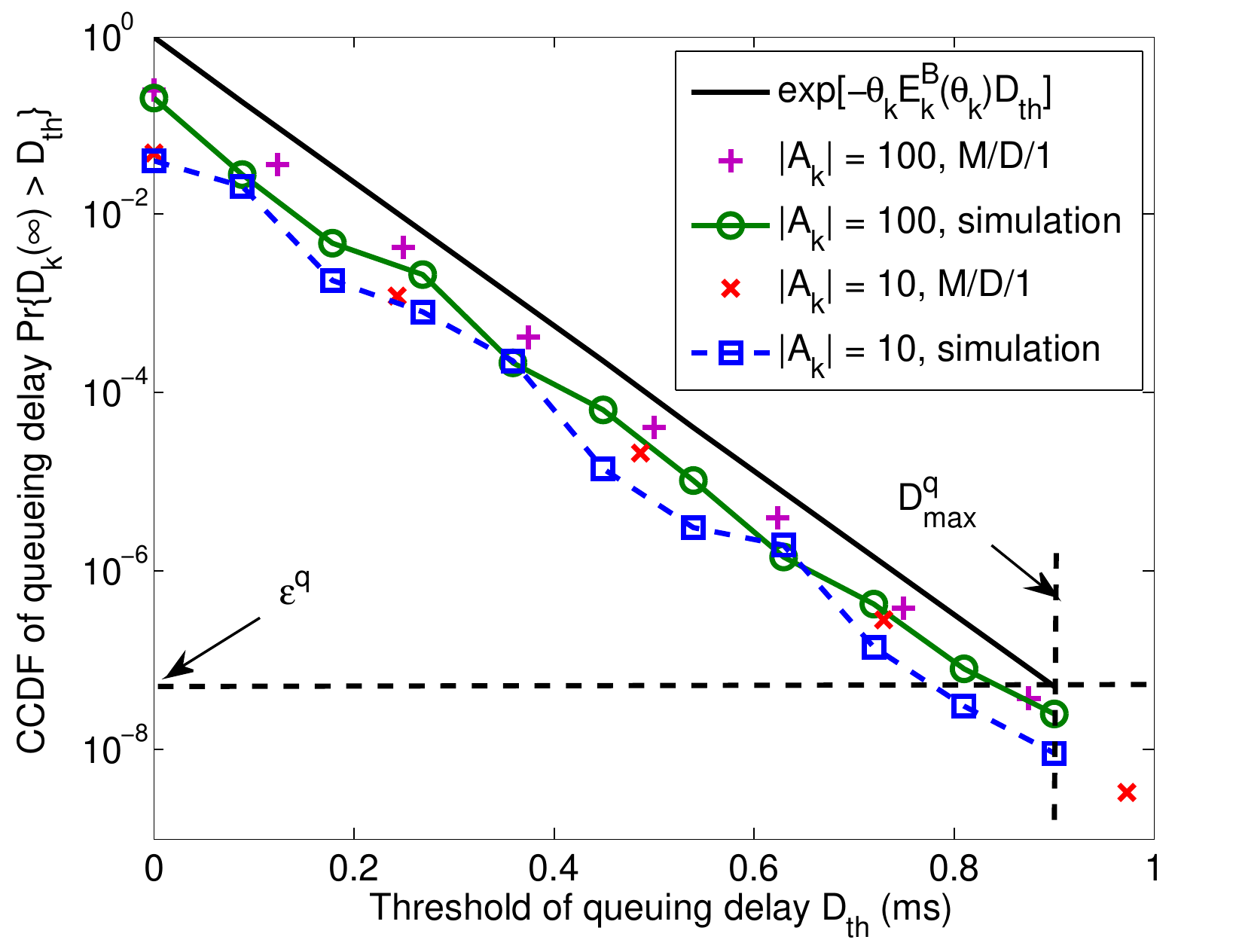}
        \end{minipage}
        \vspace{-0.3cm}
        \caption{Validating the upper bound in \eqref{eq:UB}, $N_t =8$.}
        \label{fig:CCDF}
        \vspace{-0.2cm}
\end{figure}

The CCDF of queueing delay for the packets to the $k$th user is shown in Fig. \ref{fig:CCDF}. The upper bound in \eqref{eq:UB}, i.e., $\Pr\{D_k(\infty) > D_{\rm th}\} \leq \exp\{-\theta_k E_k^B(\theta_k) D_{\rm th}\}$, is numerically obtained with different values of $D_{\rm th} \in [0,D^q_{\max}]$. The CCDF of delay with the discrete state M/D/1 model is numerically obtained from \eqref{eq:CCDFD} with $s_k(n) = E^B_k(\theta_k)$. The simulation results are obtained by computing the queueing delay of the packets served by the optimal two-state policy (the solution of problem \eqref{eq:minPtot}) during $10^{9}$ frames. It is shown that the simulated CCDFs are not smooth for short delay bounds, since the approximation in \eqref{eq:apporxD} is not accurate. However, the upper bound in \eqref{eq:UB} always exceeds the CCDFs obtained with the M/D/1 model and the simulated CCDFs, which indicates that $P_{D_k}^{\rm UB}$ is indeed an upper bound of queueing delay violation probability even when the delay bound is very short. This can be explained as follows. As shown in \eqref{eq:EBDepson}, $E^B_k(\theta_k)$ increases with $1/D^q_{\max}$. With a policy that ensures $s_k(n)=E^B_k(\theta_k)$, $s_k(n)$ also increases with $1/D^q_{\max}$. As shown in \eqref{eq:pdfMD1}, $\pi_0$ increases with $s_k(n)$. As a result, $\eta_k = 1- \pi_0$  decreases with $s_k(n)$ and hence increases with $D_{\max}^q$. When $D_{\max}^q$ is small, $\eta_k \ll 1$ (around $0.1$ in the scenario of Fig. \ref{fig:CCDF}), which leads to a loose upper bound. This indicates that the QoS constraint derived from the upper bound is conservative.

\begin{table}[htbp]
\vspace{-0.3cm}
\small
\renewcommand{\arraystretch}{1.3}
\caption{Average power consumption with finite $N_t$}
\begin{center}\vspace{-0.3cm}
\begin{tabular}{|p{3cm}|p{0.6cm}|p{0.6cm}|p{0.6cm}|p{0.6cm}|p{0.6cm}|}
  \hline
  $N_t$  & 2 & 4 & 8 & 16 & 32  \\\hline
  Normalized ${\mathbb{E}}[P_{\rm tot}(n)]$ & 1.149 & 1.042 & 1.015 & 1.005 & 1.002  \\\hline
  Normalized $P_{\rm req}^{t}$ & 0.983 & 0.482 & 0.458 & 0.442 & 0.436\\\hline
  Normalized $W_{\rm req}$ & 0.463 & 0.424 & 0.419 & 0.414 & 0.412\\\hline
\end{tabular}
\end{center}
\vspace{-0.4cm}
\end{table}

To validate that the results obtained for large value of $N_t$ are also true for not-so-large $N_t$, in Table II we provide the simulation results of the average total power consumption, required transmit power and required bandwidth with finite $N_t$, normalized by those obtained with $N_t \to \infty$ in \eqref{eq:ExpPtot}, \eqref{eq:constP} and \eqref{eq:constW}, respectively. To obtain the results, we solve problem \eqref{eq:minPtot} in $2\times 10^{6}$ frames (i.e., $10^5$ channel fading blocks) and then compute the averaged total power consumption, the maximal required transmit power and bandwidth. We set $d_u = 15$~m, $K = 160$ and $\left|{\mathcal{A}}_k\right| = 80$ for the simulation.

The results in Table II show that the average power consumption when $N_t > 2 $ is close to the lower bound in \eqref{eq:ExpPtot}. This indicates that the two-state policy is nearly EE-optimal, despite that the introduced QoS constraint is conservative. We can observe that the required maximal transmit power and bandwidth are less than the upper bounds in \eqref{eq:constP} and \eqref{eq:constW}. This is because the upper bounds are obtained under the assumption that all the buffers are not empty. However, as show in Fig. \ref{fig:CCDF}, there is very high probability that a buffer is empty. When $K$ is large, the number of users that have non-empty buffers is much less than $K$. Therefore, the required total transmit power and bandwidth is less than the upper bound.

\begin{figure}[htbp]
        \vspace{-0.3cm}
        \centering
        \begin{minipage}[t]{0.38\textwidth}
        \includegraphics[width=1\textwidth]{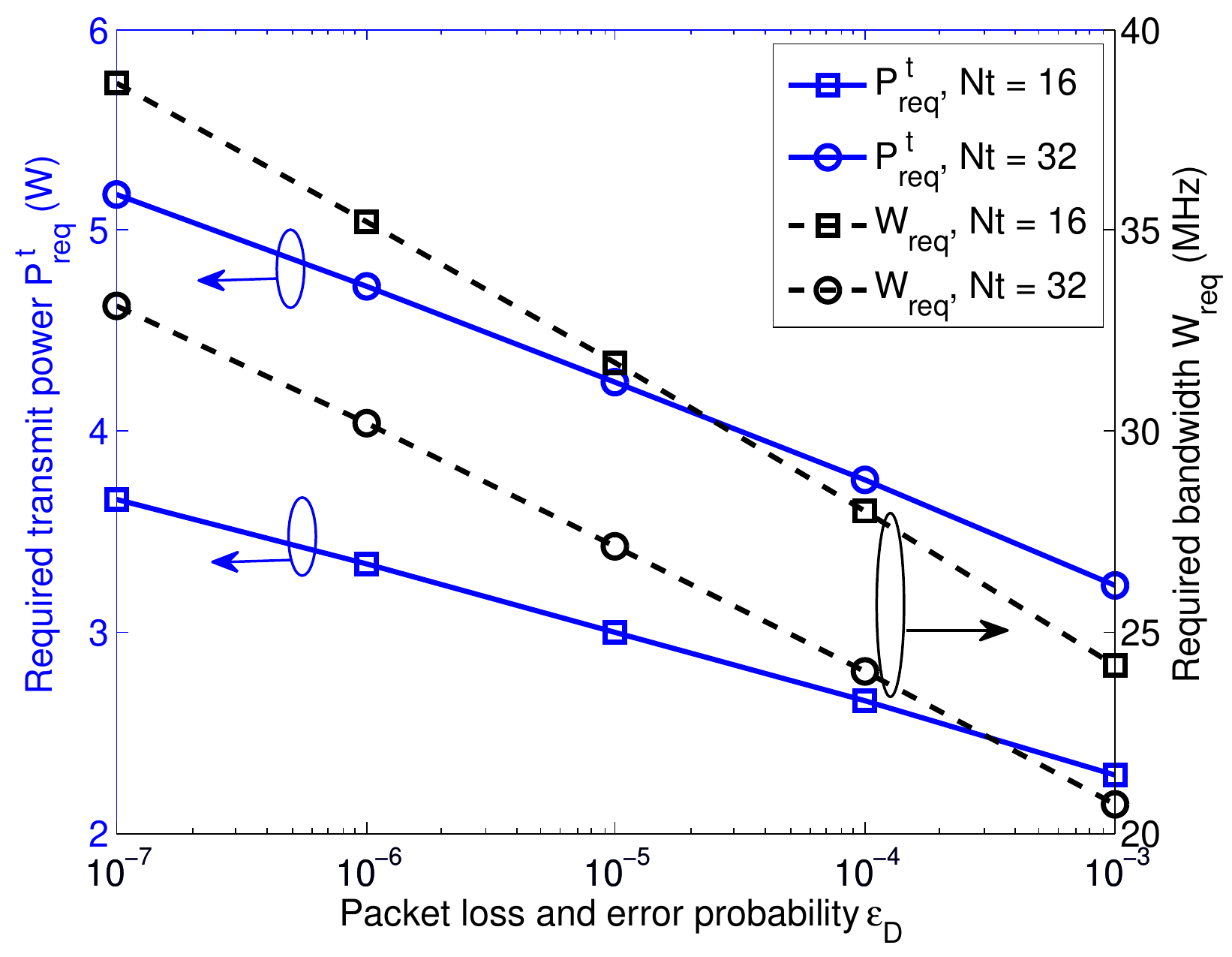}
        \end{minipage}
        \vspace{-0.3cm}
        \caption{Required maximal transmit power and bandwidth vs. reliability, $D_{\max} = 1$~ms $, d_u = 15$~m, $K = 160$ and $\left|{\mathcal{A}}_k\right| = 80$. }
        \label{fig:Req}
        \vspace{-0.3cm}
\end{figure}
In Fig. \ref{fig:Req}, we provide the upper bounds of the maximal transmit power and bandwidth required to achieve the EE limit with guaranteed QoS, which are  numerically obtained from the right-hand side of \eqref{eq:constP} and \eqref{eq:constW}. The results show that the required resources linearly increases with $\ln{(1/\varepsilon _D)}$. This means that approaching the EE limit under the ultra-high reliability and ultra-low latency  requirement does not need high transmit power or large bandwidth. We can see that the required bandwidth decreases with $N_t$, but the required transmit power increases with $N_t$. This can be explained as follows. Since $P^{cw}$ increases with $N_t$, less bandwidth should be used to reduce circuit power consumption when $N_t$ is large. With less bandwidth, more transmit power is required  to ensure QoS.

\section{Conclusion}
In this paper, we studied how to design energy efficient resource allocation in tactile internet. To ensure the delay bound and its violation probability, an upper bound of the CCDF of queueing delay derived based on the effective bandwidth was applied. We optimized a QSI and CSI dependent resource allocation policy to maximize the EE under the QoS constraint. We then showed that the minimal average total power consumption achieved by the optimized policy under the strict delay requirement equals to that under the infinite queueing delay requirement with large number of transmit antennas, which implies that the policy is optimal in maximizing the EE. Simulation and numerical results validated our analysis and showed that the achieved EE of the proposed resource allocation policy is closed to the upper bound of EE even for small number of antennas.


\bibliographystyle{IEEEtran}
\bibliography{ref}

\begin{thebibliography}{10}
\providecommand{\url}[1]{#1}
\csname url@samestyle\endcsname
\providecommand{\newblock}{\relax}
\providecommand{\bibinfo}[2]{#2}
\providecommand{\BIBentrySTDinterwordspacing}{\spaceskip=0pt\relax}
\providecommand{\BIBentryALTinterwordstretchfactor}{4}
\providecommand{\BIBentryALTinterwordspacing}{\spaceskip=\fontdimen2\font plus
\BIBentryALTinterwordstretchfactor\fontdimen3\font minus
  \fontdimen4\font\relax}
\providecommand{\BIBforeignlanguage}[2]{{%
\expandafter\ifx\csname l@#1\endcsname\relax
\typeout{** WARNING: IEEEtran.bst: No hyphenation pattern has been}%
\typeout{** loaded for the language `#1'. Using the pattern for}%
\typeout{** the default language instead.}%
\else
\language=\csname l@#1\endcsname
\fi
#2}}
\providecommand{\BIBdecl}{\relax}
\BIBdecl

\bibitem{Gerhard2014The}
G.~P. Fettweis, ``The tactile internet: Applications \& challenges,''
  \emph{IEEE Vehic. Tech. Mag.}, vol.~9, no.~1, pp. 64 -- 70, Mar. 2014.

\bibitem{A2014Scenarios}
{A. Osseiran, F. Boccardi and V. Braun, \emph{et al.}}, ``Scenarios for 5{G}
  mobile and wireless communications: The vision of the {METIS} project,''
  \emph{IEEE Commun. Mag}, vol.~52, no.~5, pp. 26 -- 35, May. 2014.

\bibitem{Changyang2016TVC}
C.~She and C.~Yang, ``Ensuring the quality-of-service of tactile internet,'' in
  \emph{Proc. IEEE VTC Spring}, 2016.

\bibitem{Petteri2015A}
{P. Kela and J. Turkka, \emph{et al.}}, ``A novel radio frame structure for
  5{G} dense outdoor radio access networks,'' in \emph{Proc. IEEE VTC Spring},
  2015.

\bibitem{Kai2014Polar}
K.~Niu, K.~Chen, J.~Lin, and Q.~T. Zhang, ``Polar codes: Primary concepts and
  practical decoding algorithms,'' \emph{IEEE Commun. Mag}, vol.~52, no.~7, pp.
  192--203, Jul. 2014.

\bibitem{YangGR2015}
G.~Wu, C.~Yang, S.~Li, and G.~Li, ``Recent advance in energy-efficient networks
  and its application in {5G} systems,'' \emph{IEEE Wireless Commun. Mag.},
  vol.~22, no.~2, pp. 145 -- 151, Apr. 2015.

\bibitem{EELIU}
L.~Liu, Y.~Yi, C.~J.-F., and J.~Zhang, ``Energy-efficient power allocation for
  delay-sensitive multimedia traffic over wireless systems,'' \emph{IEEE Trans.
  Veh. Technol.}, vol.~63, no.~5, pp. 2038 -- 2047, Mar. 2014.

\bibitem{She2015Tcom}
C.~She, C.~Yang, and L.~Liu, ``Energy-efficient resource allocation for
  {MIMO}-{OFDM} systems serving random sources with statistical {Q}o{S}
  requirement,'' \emph{IEEE Trans. Commun.}, vol.~63, no.~11, pp. 4125--4141,
  Nov. 2015.

\bibitem{EB}
C.~Chang and J.~A. Thomas, ``Effective bandwidth in high-speed digital
  networks,'' \emph{IEEE J. Sel. Areas Commun.}, vol.~13, no.~6, pp.
  1091--1100, Aug. 1995.

\bibitem{squeezing1996}
G.~L. Choudhury, D.~M. Lucantoni, and W.~Whitt, ``Squeezing the most out of
  {ATM},'' \emph{IEEE Trans. Commun.}, vol.~44, no.~2, pp. 203--217, Feb. 1996.

\bibitem{Mehdi2013Performance}
M.~Khabazian, S.~Aissa, and M.~Mehmet-Ali, ``Performance modeling of safety
  messages broadcast in vehicular ad hoc networks,'' \emph{IEEE Trans. Intell.
  Transp. Syst.}, vol.~14, no.~1, pp. 380 -- 387, Mar. 2013.

\bibitem{Yury2010Channel}
Y.~Polyanskiy, H.~V. Poor, and S.~Verd\'{u}, ``Channel coding rate in the
  finite blocklength regime,'' \emph{IEEE Trans. Inf. Theory}, vol.~56, no.~5,
  pp. 2307--2359, May 2010.

\bibitem{EC}
D.~Wu and R.~Negi, ``Effective capacity: A wireless link model for support of
  quality of service,'' \emph{IEEE Trans. Wireless Commun.}, vol.~2, no.~4, pp.
  630--643, Jul. 2003.

\bibitem{Gross1985MD1}
D.~Gross and C.~Harris, \emph{Fundamentals of Queueing Theory}.\hskip 1em plus
  0.5em minus 0.4em\relax Wiley, 1985.

\bibitem{earth2010}
\BIBentryALTinterwordspacing
{G. Auer, O. Blume, V. Giannini, I. G\'{o}dor, \emph{et al.}}, ``D 2.3: Energy
  efficiency analysis of the reference systems, areas of improvements and
  target breakdown,'' \emph{EARTH}, Jan. 2012. [Online]. Available:
  \url{https://www.ict-earth.eu/publications/deliverables/deliverables.html}
\BIBentrySTDinterwordspacing

\bibitem{Bjorn2015A}
B.~Debaillie, C.~Desset, and F.~Louagie, ``A flexible and future-proof power
  model for cellular base stations,'' in \emph{Proc. IEEE VTC Spring}, 2015.

\bibitem{boyd}
S.~Boyd and L.~Vandanberghe, \emph{{C}onvex {O}ptimization}.\hskip 1em plus
  0.5em minus 0.4em\relax Cambridge Univ. Press, 2004.

\bibitem{Rusek2013Scaling}
F.~Rusek, D.~Persson, B.~K. Lau, E.~G. Larsson, T.~L. Marzetta, O.~Edfors, and
  F.~Tufvesson, ``Scaling up {MIMO}: Opportunities and challenges with very
  large arrays,'' \emph{IEEE Signal Process. Mag.}, vol.~30, no.~1, pp. 40 --
  60, Jan. 2013.

\bibitem{She2015Optimal}
C.~She and C.~Yang, ``Optimal {EE}-delay relation in wireless systems,'' in
  \emph{Proc. IEEE Online GreenComm}, Nov. 2015.

\bibitem{Tony2015Delay}
G.~Zhang, T.~Q.~S. Quek, A.~Huang, M.~Kountouris, and H.~Shan, ``Delay modeling
  for heterogeneous backhaul technologies,'' in \emph{Proc. IEEE VTC Fall},
  2015.

\end{thebibliography}

\end{document}